# Spectroscopic evidence for spin-polarized edge states in graphitic Si nanowires

P.C. Snijders[1], P.S. Johnson[2], N.P. Guisinger[3], S.C. Erwin[4] and F.J. Himpsel[2]
[1] Materials Science and Technology Division, Oak Ridge National Laboratory, Oak Ridge, TN, USA
[2] Department of Physics, University of Wisconsin, Madison, WI, USA
[3] Center for Nanoscale Materials, Argonne National Laboratory, Argonne, IL, USA
[4] Center for Computational Materials Science, Naval Research Laboratory, Washington, DC, USA

**Abstract**
The step edges on the Si(553)–Au surface undergo a 1 × 3 reconstruction at low temperature which has recently been interpreted theoretically as the ×3 ordering of spin-polarized silicon atoms at the edges of the graphitic Si nanowires on this vicinal surface. This predicted magnetic ground state has a clear spectroscopic signature—a silicon step-edge state at 0.5 eV above the Fermi level—that arises from strong exchange splitting and hence would not occur without spin polarization. Here we report spatially resolved scanning tunneling spectroscopy data for these nanowires. At low temperature we find an unoccupied state at 0.5 eV above every third step edge silicon atom, in excellent agreement with the spin-polarized ground state predicted theoretically. This spin-polarized state survives up to room temperature where the position of the spins rapidly fluctuates among all Si step-edge sites.

**Introduction**
Dangling bonds in covalent solids can be occupied by zero, one, or two electrons. When occupied by one electron, a local magnetic moment of one Bohr magneton is associated with the dangling bond and magnetism may develop if these local moments interact, as recently reported for defects in highly oriented pyrolytic graphite and graphene nanoribbon edge states [1–3]. Notwithstanding promising efforts with self-assembled graphene nanoribbons [3, 4], current materials control does not allow the creation of large, ordered arrangements of dangling bonds carrying such local moments. Periodic surface reconstructions, on the other hand, sometimes naturally consist of exceedingly well-ordered two- or even one-dimensional arrays of partially filled broken bonds [5–8]. Such surfaces have revealed fascinating properties such as charge density waves [9–11], Mott insulating states [12, 13], and the Luttinger liquid [14]. So far, however, the presence of magnetic order arising within arrays of dangling-bond local moments has not been observed experimentally.
Exactly this type of ordering was recently predicted, on the basis of density-functional theory (DFT) calculations, to arise naturally (at T = 0) on the Au-induced reconstruction of the Si(553) surface [15]. Specifically, Erwin and Himpsel [15] predicted that the unit cell of Si(553)–Au has a terrace-plus-step structure in which the step-edge consists of a graphitic honeycomb chain of Si atoms, akin to silicene [16, 17]. Every third Si atom along the edge is fully spin-polarized, as shown in figure 1(a), while the two intervening Si atoms are nonspin-polarized. The spin–spin interactions are predominantly antiferromagnetic, but even in the absence of magnetic order the Si dangling-bond local moments are expected to survive up to room temperature.
Scanning tunneling microscopy (STM) simulations based on this structural model reveal ×3 topography along the step edge even for STM imaging without spin-resolution [15] consistent with experimental observations at low temperature [18, 19]. In empty-state STM images the spin-polarized Si atoms appear higher than the non-spin-polarized atoms, despite the fact that their actual position is lower. In the model this occurs because the spin-polarized Si atoms have a minority-spin state 0.5 eV above the Fermi level, which increases the STM tunneling current into empty states. Spectroscopic verification of such a state at the predicted step-edge location is however lacking. It was also suggested in [15] that, at room temperature, the experimentally observed ×1 periodicity could be understood as due to thermal fluctuations in the positions of the Si step-edge atoms that cannot be resolved on the STM time scale [18, 19].

Here we report a scanning tunneling spectroscopic (STS) study of the electronic structure at this surface, comparing it to the predicted electronic structure. While direct proof of a magnetic ground state of this system would require low-temperature spin-resolved STM and STS experiments, we offer here important evidence from spin-integrated measurements of the local density of states (LDOS) on the edges of the graphitic Si nanowire above and below the transition. Our results reveal the presence of an edge state that is only present in the calculated band structure when spin-polarization is included [15]. Moreover, they appear to confirm the suggestion that thermal fluctuations eventually smear out the threefold periodicity of the ground state, leading to an 'order–disorder' transition between the low-temperature ×3 ordered phase and the room-temperature '×1' disordered phase. This transition involves strongly coupled atomic and spin degrees of freedom. The spin-polarization survives at least partially up to room temperature.

**Experimental methods**
Arrays of Si step-edge dangling bonds were fabricated through self-assembly using a high-index Si surface. In particular, ordered arrays of step edges were stabilized on the surface of vicinal Si(553) by deposition of 0.46 monolayer (ML) Au, see figure 1(a). The procedure has been described previously [20] and consists of depositing Au from a heated Mo filament at a rate of ∼0.01 ML s−1 onto the sample surface held at 975 K. A post-anneal to 1120 K followed by slow cool-down to room temperature was used to produce an ordered array of atomic steps and terraces at the surface. The sample temperature during sample preparation was measured using an infrared pyrometer with the emissivity set to 0.4. STM and STS experiments were performed at sample temperatures of ∼40 and 300 K in an Omicron variable temperature STM using PtIr and W etched tips. Since the variable temperature microscope suffers from finite thermal drift, we employed a number of strategies to ensure this did not affect our spatially resolved spectroscopic results. We measured sets of five $I(V)$ and $dI/dV$ curves using 'V-shaped' (i.e. forward and backward) voltage ramps, and only accepted data for which both forward and backward directions resulted in identical data for all five of the spectra, as this indicates the tip did not drift to a different (x,y)-position during the recording of the spectra. The $(dI/dV)$ curves were then normalized by $I/V$ to $(dI/dV)/(I/V)$ curves, minimizing effects of drift and hysteresis in the z-direction.

**Theoretical methods**
First-principles calculations of the atomic, electronic, and magnetic structure of Si(553)–Au were performed using a simplified 1×3 ferromagnetic variant, shown in figure 1(a), of the full 1×6 antiferromagnetic structural model proposed in [15]. The calculations were performed in a slab geometry with six or more layers of Si plus the reconstructed top surface layer and a vacuum region of 10 Å. All atomic positions were relaxed, except the bottom Si layer and its passivating hydrogen layer, until the largest force component on every atom was below 0.02 eV Å−1. Total energies and forces were calculated within the Perdew–Burke–Ernzerhof (PBE) generalized-gradient approximation [21] to DFT using projector-augmented-wave potentials, as implemented in vasp [22, 23]. The plane-wave cutoff was 250 eV.

It is important to note that the electronic states arising from Si dangling-bond orbitals at the Si(553) step edge are quite narrow (less than 0.5 eV). As a result these states may be subject to electron correlation effects not accurately treated within PBE. To address this issue we improved our first-principles calculations by using the screened hybrid functional of HSE [24, 25] to calculate the electronic band structure. The HSE functional replaces a fraction of the PBE exchange potential by the exact Hartree–Fock exchange potential. Because we are especially interested in states related to Si dangling bonds we used a reduced mixing fraction, a = 0.11, previously shown to give an accurate description of the dangling bond in bulk Si [26].

**Results and discussion**
The DFT/HSE theoretical band structure for Si(553)–Au (figure 1(b)) shows that the predicted spin-polarized ground state has a very clear spectroscopic signature, even for non-spin-polarized measurements. This is because spin polarization leads to a large exchange splitting, more than 0.5 eV, of the Si step-edge state marked by red diamonds. Hence it predicts the existence of an unoccupied minority-spin state localized above every third step-edge Si atom, centered at 0.5 eV above the Fermi level. In a non-spin-polarized calculation, in contrast, the corresponding state has equal weight on every step-edge Si atom and never extends higher than 0.1 eV above the Fermi level [15]. This difference allows for a direct test, using standard spectroscopic approaches, of the predicted spin-polarized ground state. Note that for occupied states the distinction between spin-polarized and non-spin-polarized configurations is less striking. Hence we will focus on the unoccupied LDOS in the STS data below.

Figure 2 presents empty-state STM images of the Si(553)–Au surface at room temperature and 40 K. The bright ridges visible in the images originate from the orbitals at the edge of the strip of graphitic silicon atoms while the dark valleys are located at the double row of Au atoms on the terrace in each unit cell of the stepped surface. At 40 K the topography shows a clear ×3 periodicity with one bright and two dark atoms at the step-edge, as reported previously [18, 19].

Figure 3 presents two normalized STS spectra, $(dI/dV)/(I/V)$, measured on the ×1 topography of Si(553)–Au at room temperature. These spectra are proportional to the LDOS of the sample at the location where the spectra are taken. The two spectra were measured on the Si step edge and on the Au chain, respectively (see figure 2(a)). Consistent with earlier reports [27, 28], the LDOS at the Si step edge exhibits a state at 0.5 V. Figure 3 shows that this state is absent at the location of the Au atoms. The peak at 0.25 V is present at both the Si step edge and the Au chain and is therefore not a step-edge state. Apart from a difference in amplitude, the LDOS of the occupied states is qualitatively similar for the two locations.

At low temperatures, spectra measured on the Au chain (not shown) are similar to those at room temperature. But figure 4 shows that spectra measured on the bright and dark Si atoms of the step edge reveal a striking difference compared to room temperature: the peak at 0.5 V now occurs only on the bright atoms, and is larger in amplitude. This result is in excellent agreement with the theoretical prediction that at low temperature every third Si atom is spin polarized with an unoccupied minority state at 0.5 eV. It is also supported by recent two-photon photoemission experiments probing the unoccupied electronic states [29], which are inaccessible to conventional photoemission [18, 20, 30]. Figure 4 shows that for the occupied states, the spectra on bright and dark atoms are indistinguishable. This is also in good agreement with the theoretical predictions.

At low temperatures, the peak at 0.25 V is present on both the bright (polarized) and the dark (non-polarized) Si step-edge atoms, as well as the Au chain atoms (see figure 3), and thus likely does not originate from spin-polarized parts of the electronic structure. The origin of this peak is therefore not clear at this moment, although it is interesting to note that the bottom of the empty minority spin state has a significant weight at lower energies, and could hybridize with the inner Au band, see figure 1.

At room temperature, the apparent periodicity of the STM topography along the step edge is ×1. In [15] this observation was tentatively attributed to rapid local fluctuations of Si atoms between spin-polarized and non-spin-polarized states. Thus the system exhibits an order-disorder transition above a temperature that we now estimate, from preliminary molecular dynamics simulations, to be in the range 30–70 K. These molecular-dynamics simulations of this transition will be the subject of a forthcoming publication. The experimental signature of this disordered state would be the appearance of a spin-polarized empty state on all Si step-edge atoms, but with a smaller amplitude than in the low-temperature state. This is indeed what we observe in the STS spectra of figure 3: the peak at 0.5 eV detected at low temperature only on the bright Si atoms is still present in the room-temperature spectra, but with a smaller amplitude and on every Si step-edge atom. The presence of an order–disorder transition temperature in the range mentioned above is consistent with temperature-dependent STM imaging reported in [19] showing a coexistence of the ordered and disordered phases between 40 and 110 K, and the finite correlation length of the ordered phase at 47 K even in long chains reported in [34].

**Summary**
We have shown that the unoccupied states at the edge of the graphitic Si nanowire on the Si(553)–Au surface are the key to confirming the spin-polarized ground state predicted theoretically. Localized tunneling spectra reveal a large difference in the unoccupied LDOS at the Si step edge compared to the Au chains—establishing the existence of a state localized on the step-edge atoms. At low temperature, every third atom along the step edge gives rise to a strong peak in the tunneling spectra 0.5 eV above the Fermi level—in excellent agreement with the large exchange splitting predicted theoretically for every third Si step-edge atom. At room temperature, this peak appears on every atom along the step edge, consistent with an order–disorder transition due to thermal excitation.

The perfectly linear chains of spins reported in our work offer unique opportunities for designing an atomic scale memory or information processing. In fact, a spin-shift register has already been proposed recently [31], and the self-assembled spin chains analyzed in the current work provide a first system in which prototypes of this approach could possibly be realized experimentally. The capability of reading and writing individual spins in nanostructures has recently been demonstrated by the IBM Almaden group [32] and the Wiesendanger group at Hamburg [33].

**Acknowledgments**

Part of this research was supported by the US Department of Energy, Basic Energy Sciences, Materials Sciences and Engineering Division (PCS). FJH and PSJ acknowledge support by the NSF under Award No. DMR-0705145. Part of this work was supported by the Office of Naval Research (SCE). The DFT computations were performed at the DoD Major Shared Resource Centers at AFRL and ERDC. Use of the Center for Nanoscale Materials was supported by the US Department of Energy, Office of Science, Office of Basic Energy Sciences, under Contract No. DE-AC02-06CH11357. We thank H H Weitering for kindly providing the Si(553) wafer.

Figure 1

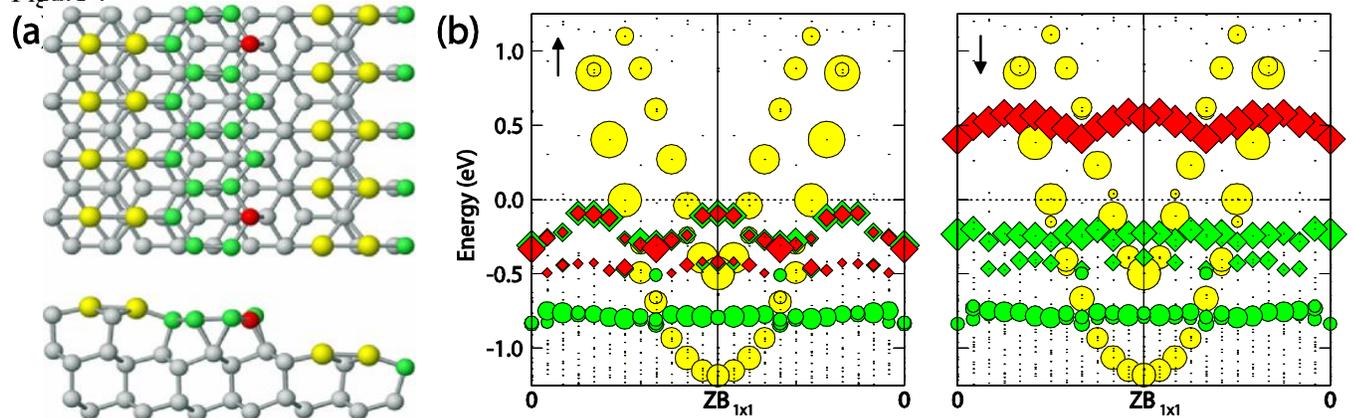

*Figure 1.* (a) Top and side views of the Si(553)–Au structural model. Large yellow atoms are Au, all others are Si, with the graphitic Si nanowire indicated with green and red atoms. The periodicity along the step edge is ×3 due to the predicted 3a spacing of spin-polarized Si atoms (red). (b) Theoretical DFT/HSE (Heyd, Scuseria, and Ernzerhof) band structure of Si(553)–Au. Red diamonds are states from spin-polarized Si atoms at the step edge. The red spin-down (minority) state at 0.5 eV is created by magnetic exchange splitting and hence does not occur without spin polarization. Yellow circles are states from Au, green diamonds from non-polarized Si, and green circles from non-polarized Si at the opposite edge of the green graphitic Si nanowire.

Figure 2

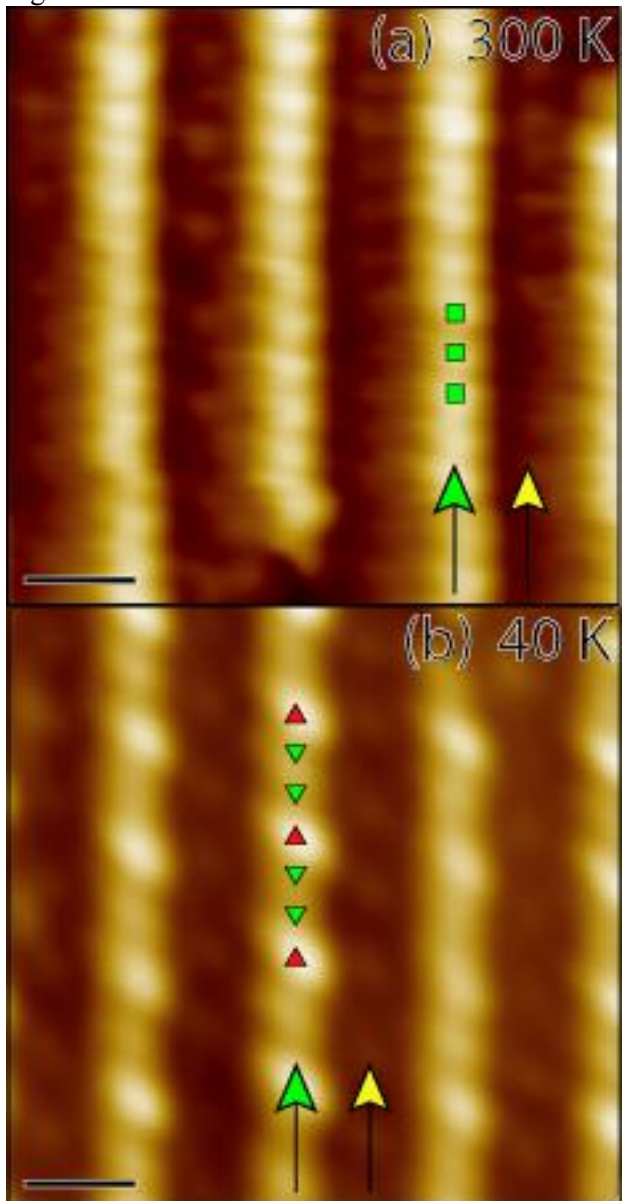

*Figure 2*. STM images of the Si(553)–Au surface at (a) room temperature, and (b) 40 K showing 1 × 1 and 1 × 3 periodicities. The edge of the strip of graphitic Si and the Au chain are indicated with green and yellow arrows, respectively. At 40 K, the spin-polarized Si edge atoms (red upward triangles) appear to be higher than non-spin-polarized atoms (green downward triangles) in this empty-state topography because of their associated unoccupied minority-spin state (see text). Scale bars: 1 nm. V = 0.5 V (a) and V = 1 V (b), I = 100 pA.

Figure 3

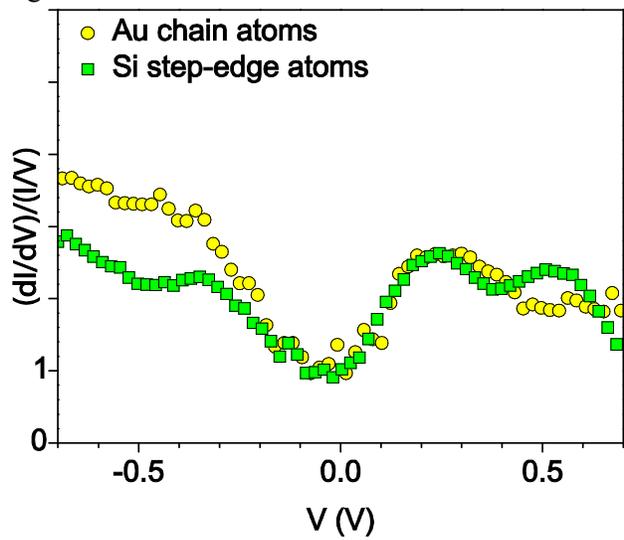

*Figure 3*. Normalized STS $(dI/dV)/(I/V)$ spectra taken at room temperature on the edge of the graphitic Si strip and the Au chain at the center of the terrace, see figure 2(a). The peak at +0.5 V exists only at the step edge.

Figure 4

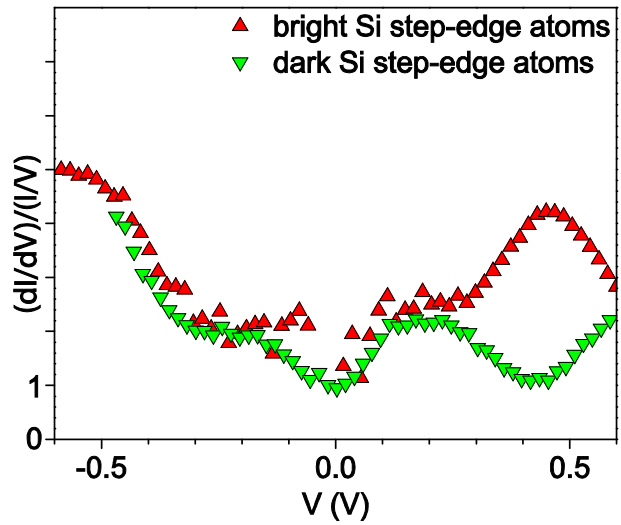

*Figure 4*. Normalized STS (d I /dV )/( I / V ) spectra taken at 40 K on the bright and dark Si step-edge atoms, see figure 2(b). A clear difference is evident in the unoccupied states, for which the bright atoms exhibit a strong peak at 0.5 eV not seen on the dark atoms. This peak is in excellent agreement with the predicted unoccupied minority-spin state of the spin-polarized Si atoms, shown in figure 1(b).